 \documentstyle[preprint,epsf,aps,amsfonts]{revtex}
\newcommand{\be}{\begin{equation}}
\newcommand{\ee}{\end{equation}}
\newcommand{\p}{\prime}
\newcommand{\al}{\alpha}
\newcommand{\ba}{\beta}
\newcommand{\de}{\delta}

\newcommand{\r}{\rho}
\newcommand{\g}{\gamma}
\newcommand{\ov}{\over}
\newcommand{\th}{\theta}
\newcommand{\ve}{\varepsilon}
\renewcommand{\thefootnote}{\fnsymbol{footnote}}

\begin{document}

%%\begin{titlepage}
{\tighten
\preprint{\vbox{\hbox{CALT-68-1998}
               \hbox{USC-95-015}
               \hbox{hep-th/9506067}
               \hbox{\footnotesize DOE RESEARCH AND}
               \hbox{\footnotesize DEVELOPMENT REPORT} }}

 \title{Surface excitations and surface energy
 of the  antiferromagnetic XXZ chain by the Bethe ansatz approach.}

 \vspace{4mm}

 \author{A.Kapustin\footnote[2]{Work supported in part by
 the U.S. Dept. of Energy under Grant no. DE-FG03-92-ER40701.}}

\address{Physics Department, California Institute
 of Technology, Pasadena, CA 91125}

\author{S.Skorik\footnote[1]{Work supported by
 Grant NSF-PHY-9357207.}}

\address{Physics Department,
University of Southern California, Los Angeles, CA 90089-0484}

\maketitle

 \begin{abstract}
We study boundary bound states using the Bethe ansatz formalism
for the finite $XXZ$ $(\Delta>1)$
 chain in a boundary magnetic field $h$.
 Boundary bound states are represented by
the ``boundary strings'' similar to those described in \cite{SS}. We find
that for certain values of $h$
 the ground state wave function contains
 boundary strings, and from this infer the existence of two ``critical''
fields in agreement with \cite{MJ}.
 An expression
for the vacuum surface energy in the thermodynamic limit is derived and found
to
be an analytic function of $h$.
We argue that boundary excitations appear only in pairs
with ``bulk'' excitations or with boundary excitations at the other end of the
chain. We mainly discuss the case where
the magnetic fields at the left and the right boundaries are antiparallel, but
we also comment on the case of the parallel fields.
In  the Ising ($\Delta=\infty$) and isotropic ($\Delta=1$)
limits our results agree with those previously known.

 \end{abstract}
}
%% \end{titlepage}
\newpage

\renewcommand{\thefootnote}{\arabic{footnote}}

%% \vspace{2mm}

One-dimensional integrable
quantum field theories with boundary interactions \cite{GZ} have been
intensively studied recently because of their applications in condensed matter
physics (see e.g.,\cite{CMP}). A powerful method for dealing with such
problems
is the Bethe ansatz, allowing basic physical properties to be extracted
from the
system of coupled transcendental equations. Among others, it allows to
solve the boundary sine-Gordon model via its lattice regularization, the
inhomogeneous $XXZ$ $(|\Delta|<1)$
 chain in a boundary magnetic field  \cite{SS},\cite{FS}.

In this Letter we study
the $XXZ$ chain with even number of spins $L$ in a boundary magnetic field,
\be
{\cal H}= {1\over 2 }
\left\{\sum_{i=1}^{L-1} \left(
\sigma^x_i \sigma^x_{i+1} + \sigma^y_i
\sigma^y_{i+1}+\Delta\sigma^z_i
\sigma^z_{i+1}\right) +h_1\sigma_1^z+h_2\sigma_L^z \right\}, \label{ham}
\ee
in the regime $\Delta>1$, $h_1\geq 0$, $h_2\leq 0$,
 focusing on the effects peculiar to systems with boundaries
\cite{MW}. At $h_1=h_2=0$ this model describes one-dimensional antiferromagnet
 with non-magnetic impurities, accessible experimentally. We exploit the Bethe
 ansatz
solution for this model, first derived  in \cite{ABBBQ}, together with the
well-known results for the periodic chain \cite{GAU}. We find new ``boundary
string'' solutions to the Bethe equations, similar to the boundary strings
existing in the $|\Delta|<1$ regime \cite{SS}. For certain
 values of the boundary
magnetic field the ground state configuration
contains boundary 1-strings. Boundary excitations are obtained by removing
(or adding, depending on the sign of $h$) boundary strings from the
 ground state wave-function.
 Their energy was first obtained in \cite{MJ} by the algebraic approach.

A peculiar feature of the Bethe ansatz solution of the periodic chain is that
 the excitations (holes in the Dirac sea) appear only in pairs
\cite{FT}. We argue that similarly the boundary excitations
can appear only in pairs with bulk excitations or with boundary excitations at
the other end of the spin chain. This is not evident in the algebraic approach
 of \cite{MJ}.

Using  the Bethe ansatz
solution we calculate the surface energy (see e.g.,\cite{BH}):
\be
E_{surf}(L,\Delta, h)=E_{gr}-E^0_{gr}, \label{surfen}
\ee
 in the thermodynamic limit $L=\infty$. Here $E_{gr}$
is the ground state energy of (\ref{ham}) and  $E_{gr}^0$ is that of the
periodic
chain. We give an interpretation of our results in the
limits $\Delta\to\infty$ and $\Delta\to 1$, corresponding to the 1D Ising
and XXX models respectively. Finally, we comment on the structure of the
ground state when the boundary magnetic fields are parallel.

Let us first set up the Bethe ansatz (BA) notations and list the relevant
 results about the XXZ chain \cite{ABBBQ,GAU}.
In \cite{ABBBQ} the eigenstates of (\ref{ham}) were constructed
for arbitrary $\Delta$.
 As usual in the BA picture, the $n$-magnon eigenstates $|n\rangle$, satisfying
${\cal H}|n\rangle=E|n\rangle$,
are linear combinations of
the states with $n$ spins down, located at sites
$x_1,...,x_n$:
$$|n\rangle=\sum
f^{(n)}(x_1,...,x_n)|x_1,...,x_n\rangle.
$$
The wave-function
\be
 f(x_1,...,x_n)=\sum_P \ve_P A(p_1,...,p_n)e^{i(p_1x_1+...+p_nx_n)},
\label{wf}
\ee
contains $n$ parameters $p_j\in(0,\pi)$ which are subject
to quantization conditions, called Bethe equations (BE):
\be
e^{2iLp_j}\cdot{e^{ip_j}+h_1-\Delta\over 1+(h_1-\Delta)e^{ip_j}}\cdot
{e^{ip_j}+h_2-\Delta\over 1+(h_2-\Delta)e^{ip_j}}=\prod_{l\neq j}^n
e^{i\Phi(p_j,p_l)}. \label{beq}
\ee
The energy and spin of the
$n$-magnon state are given by \cite{ABBBQ}:
\be
E={1\ov 2}[(L-1)\Delta+h_1+h_2]+2\sum_{j=1}^n(\cos p_j-\Delta),
\qquad S_z={1\ov 2}L-n. \label{ene}
\ee

It is convenient to rewrite BE using the following mappings:
\be
 \Delta=\cosh\g \geq 1, \qquad \g\geq 0, \label{mapI}
\ee
\be
p=-i\ln{\cosh{1\over 2}(i\alpha+\gamma)\ov \cosh{1\ov 2}(i\al-\g)},
\label{mapII}
\ee
(our definition of $p(\alpha)$ differs from that of \cite{GAU} by the shift
$\al\to\al+\pi$ and it was chosen in such a way that
$p(\al)$ be an odd function that  maps $-\pi<\al<\pi$ to $-\pi<p<\pi$),
\be
 h=\cosh\g +{\sinh{\g\ov 2}(1-H)\ov\sinh{\g\ov 2}(1+H)}=
\sinh\g\cdot\coth{\g\ov 2}(H+1), \qquad
h_{lim}<|h|<\infty, \label{maphI}
\ee
\be
 h=\cosh\g -{\cosh{\g\ov 2}(1-H)\ov\cosh{\g\ov 2}(1+H)}=
\sinh\g\cdot\tanh{\g\ov 2}(H+1), \qquad |h|<h_{lim}.
\label{maphII}
\ee
The latter two mappings are defined on $H\in(-\infty,\infty)$ and are
necessary to cover the region $-\infty<h<\infty$, with
positive $h$ corresponding to $H\in(-1,\infty)$.
 The value
$h_{lim}\equiv h(\infty)=\sinh\g$ lies between two critical fields
 $h_{cr}^{(1)},h_{cr}^{(2)}$ defined as follows \cite{MJ} :
\be
 h_{cr}^{(1)}=\Delta-1, \qquad h_{cr}^{(2)}=\Delta+1. \label{crfields}
\ee
Both critical fields correspond to $H=0$, and the gap $h_{cr}^{(1)}<h<
h_{cr}^{(2)}$ corresponds to $0<H<\infty$.
In these notations eq. (\ref{beq}) becomes:
\be
\left[{\cosh{1\over 2}(i\alpha_j+\g)\over
\cosh{1\over 2}(i\alpha_j-\g)}
\right]^{2L} B(\al_j,H_1) B(\al_j, H_2)
 =\prod_{ m\neq j}
 {\sinh{1\over 2}(i\alpha_j-i\alpha_m+2\g)
\sinh{1\over 2}(i\alpha_j+i\alpha_m+2\g)\over
\sinh{1\over 2}(i\alpha_j-i\alpha_m-2\g)
\sinh{1\over 2}(i\alpha_j+i\alpha_m-2\g)},   \label{be}
\ee
where
\be
 B(\al,H)={\cosh{1\over 2}(i\alpha+\g H)\over
\cosh{1\over 2}(i\alpha-\g H)}, \qquad h_{lim}<|h|<\infty, \label{defBI}
\ee
\be
 B(\al,H)={\sinh{1\over 2}(i\alpha+\g H)\over
\sinh{1\over 2}(i\alpha-\g H)}, \qquad |h|<h_{lim}, \label{defBII}
\ee
are called boundary terms.
The energy eq. (\ref{ene}) takes the form:
\be
E={1\ov 2}[(L-1)\cosh\gamma+h_1+h_2]-2\sinh\g\sum_{j=1}^n p'(\al_j), \qquad
 p'(\al)={\sinh\g\ov\cosh\g+\cos\al}.
\label{ener}
\ee

In the thermodynamic limit $L\to\infty$ the real roots $\alpha_j$ of BE
form a dense
distribution in the open interval $(0,\pi)$
with density $\rho(\al)$, $dI=2L(\rho+\r^h)d\al$ being the number of
roots in the interval $d\al$. The logarithm of eq. (\ref{be}) is:
\be
 2Lp(\al_j)+{1\ov i}\ln B(\al_j,H_1)+{1\ov i}\ln B(\al_j,H_2)+\phi(2\alpha_j)
=\sum_{l=1}^n \phi(\al_j-\al_l)+\phi(\al_j+\al_l) +
2\pi I_j, \label{lbe}
\ee
where $I_j$ form an increasing sequence of positive integers, and
\be
\phi(\al)=-i\ln{\sinh{1\ov 2}(2\g+i\al)\ov\sinh{1\ov 2}(2\g-i\al)},
\qquad \phi(0)=0. \label{phase}
\ee
Taking the derivative of eq. (\ref{be}) and defining $\rho$ for negative
$\al$ by $\rho(\al)=\rho(-\al)$, we obtain
\be
 p'(\al)+{1\ov 2L}p_{bdry}^{\p}(\al)=\int_{-\pi}^{\pi}\phi'(\al-\ba)\rho(\ba)d
\ba
+2\pi(\rho(\al)+\rho_h(\al)),  \label{inteq}
\ee
with
\be
p_{bdry}^{\p}(\al)=-i{B'(\al,H_1)\ov B(\al,H_1)}-
i{B'(\al,H_2)\ov B(\al,H_2)}+2\phi'(2\al)-
2\pi\delta(\al)-2\pi\delta(\al-\pi). \label{pbdry}
\ee
The presence of delta-functions  in (\ref{pbdry}) is due to the fact that
$\al_j=0$ and $\al_j=\pi$ are always solutions to (\ref{be}),
which should be excluded, since they make the wave-function
(\ref{wf}) vanish identically \cite{FS}.

In eq. (\ref{inteq}) the ``boundary terms''
 are down by the factor  $1/2L$. Neglecting
$p_{bdry}'$ and setting
$\rho_h=0$, we obtain the equation for the ground state density
 of the periodic XXZ chain \cite{YY}. Solving it by
the Fourier expansion
\be
f(\al)=\sum_{l=-\infty}^{\infty}\hat{f}(l)e^{il\al}, \qquad
\hat{f}(l)={1\ov 2\pi}\int_{-\pi}^{\pi}f(\al)e^{-il\al}d\al, \label{fourier}
\ee
 and using (\ref{ener}), we recover the result for the
ground state energy of the periodic chain \cite{YY}:
\be
2\pi\hat{\r}_{per}(n)={ \hat{p}'(n) \ov 1+\hat{\phi}'(n)},\qquad
\hat{\phi}'(n)=e^{-2\g|n|}, \qquad \hat{p}'(n)=(-1)^ne^{-\g|n|}
\label{useful}
\ee
\be
E_{gr}^0={L\Delta\ov 2}
-2L\sinh\g\int_{-\pi}^{\pi}\r_{per}(\al)p'(\al)d\al={L\Delta\ov 2}
-L\sinh\g\sum_{n=-\infty}^{\infty}
{e^{-\g|n|}\ov\cosh\g n}. \label{grener}
\ee
The spin of the ground state is
$S_z=L/2-L\int_{-\pi}^{\pi}\r_{per} d\al=0,$
which is the well-known result \cite{YY}.

An elementary ``bulk'' excitation above the vacuum
in the model (\ref{ham}) is a hole in the distribution
of $I_j$, but only a pair of holes
can occur for the periodic chain, as argued in \cite{FT}. Thus physical
 excitations contain an even number of holes.
%For the single hole we get the equation
%$$ p'(\al)+{1\ov 2L}p_{bdry}^{\p}(\al)=\int_{-\pi}^{\pi}\phi'(\al-\ba)
%\tilde{\r}(\ba)d\ba -{1\ov 2L}\phi'(\al-\th)-{1\ov 2L}\phi'(\al+\th)
%+2\pi\tilde{\r}(\al)$$
%Subtracting (\ref{inteq}) from the latter and denoting %$\de\r=2L(\tilde{\r}-
%\r)$
%we obtain
%$$0=\int_{-\pi}^{\pi}\phi'(\al-\ba)
%\de\r(\ba)d\ba -\phi'(\al-\th)-\phi'(\al+\th)+2\pi\de\r$$
The energy of the hole with rapidity $\theta$ can be easily computed:
\be
\ve_h(\th)=
%E-E_{gr}=2\sinh\g p'(\th)-\sinh\g\int_{-\pi}^{\pi}\de\r p'(\al)d\al=
\sinh\g\sum_{n=-\infty}^{\infty}{(-1)^ne^{in\th}\ov \cosh\g n}>0,\label{hole}
\ee
and the spin with respect to the vacuum is $S_z=
%1-\int_{0}^{\pi}\de\r=
1/2$.  (Our result, eq. (\ref{hole}), differs
from the conventional one by the shift $\th\to\th+\pi$, but the dispersion
relation is unchanged by rapidity reparametrization.)
%It is useful to list here some of the formulas we are using:
%Everywhere index $n$ under the sum means that summation is from $-\infty$
%to $+\infty$.
%$$\phi'={\sinh 2\g \ov \cosh 2\g - \cos\al}>0$$
%$$ 2\pi\hat{\de\r}_{hole}(n)={2\cos n\th e^{-2\g|n|}\ov 1+e^{-2\g|n|}}$$

Analogous arguments can be applied to analyze ``bulk''
string solutions with complex $\alpha$. Although there exists
an infinite hierarchy of complex strings of arbitrary length
and quartets, their energy vanishes with respect to the vacuum
  \cite{BVV}.

So far we discussed the bulk excitations, which are essentially
the same as in the periodic chain.
Let us now turn to the new solutions of eq. (\ref{be}), boundary strings.
The analysis is close to that of \cite{SS}. Boundary excitations
have their wave-function (\ref{wf}) localized at $x=0$ or $x=L$,
and in the limit $L\to\infty$ the two boundaries may be considered separately.
Let us study first the left boundary, $h_1>0$.
The fundamental boundary 1-string
consists of one root located at $\al_0=-i\g H_1$ for $0<h_1<h_{cr}^{(1)}$,
and at $\al_0=\pi-i\g H_1$ for $h_{cr}^{(2)}<h_1<\infty$
(in both cases $-1<H_1<0$).
It is a solution of BE due to the mutual cancellation of the decreasing
 modulus of the first term in (\ref{be}) and the increasing modulus of the
second term $B(\al, H_1)$ as $L\to\infty$ and $\al\to\al_0$.
When $h_{cr}^{(1)}<h_1<h_{cr}^{(2)}$, no such solution exists.
Introduction of such a string into the vacuum
whose root density $\rho(\al)$ satisfies
\be
 p'(\al)+{1\ov 2L}p_{bdry}^{\p}(\al)=\int_{-\pi}^{\pi}\phi'(\al-\ba)\rho(\ba)d
\ba
+2\pi\rho(\al),  \label{inteqII}
\ee
 leads to the redistribution
of roots described by $\de\r\equiv 2L(\tilde{\r}-\r)$. $\de\r$ is a solution of
the equation
\be
0=\int_{-\pi}^{\pi}\phi'(\al-\ba)
\de\r(\ba)d\ba +\phi'(\al-\al_0)+\phi'(\al+\al_0)
+2\pi\de\r, \label{densI}
\ee
Solving (\ref{densI}), we find
\be
2\pi\de\hat{\r}(n)=-{2\cos n\al_0 e^{-2\g|n|}\ov 1+e^{-2\g|n|}}.
\label{answ}
\ee
For the energy and spin of the boundary 1-string with respect to the Dirac sea
we obtain:
\be
\tilde{\ve}_b=
-2\sinh\g p'(\al_0)-\sinh\g\int_{-\pi}^{\pi}\de\r p'(\al)d\al=
-\sinh\g\sum_{n=-\infty}^{\infty}{(-1)^ne^{in\al_0}\ov\cosh\g n}, \qquad
S_z=-1/2. \label{ansb}
\ee
We see that the energy (\ref{ansb}) is negative, so the
correct ground state wave-function (\ref{wf})
should contain the boundary 1-string root $\al_0$ when the value of
$h_1$ is not in the gap $h^{(1)}_{cr}<h_1<h^{(2)}_{cr}$.
The ground state density $\tilde{\rho}$ in this case satisfies the equation
\be
 p'(\al)+{1\ov 2L}(p_{bdry}^{\p}(\al)-\phi'(\al-\al_0)-\phi'(\al+\al_0))
=\int_{-\pi}^{\pi}\phi'(\al-\ba)
\tilde{\rho}(\ba)d\ba
+2\pi\tilde{\rho}(\al).  \label{inteqI}
\ee
 The boundary excitation
is obtained by removing from vacuum the root $\al_0$, which means
that it has the energy $-\tilde{\ve}_b>0$
and spin $1/2$, equal to the spin
of the bulk elementary excitation. Substituting the value of $\al_0$ into
 (\ref{ansb}), we
get the boundary excitation energy, which precisely agrees with
the one obtained in \cite{MJ}:
\be
\ve_b(h_1)=\sinh\g\sum_{n=-\infty}^{\infty}{(-1)^{\kappa n}e^{\g H_1n}
\ov\cosh\g n},
\qquad -1<H_1<0,
\label{ansen}
\ee
with $\kappa=1$ if $h_1<h_{cr}^{(1)}$ and $\kappa=2$ if $h_1>h_{cr}^{(2)}$.

{}From (\ref{hole}) and (\ref{ansen}) we see that for $h_1<h_{cr}^{(1)}$ the
 energy of the boundary
excitation is smaller than the energy of the zero-rapidity bulk excitation, and
becomes equal to it at $h_1=h_{cr}^{(1)}$. So in this regime we may interpret
 the boundary excitation as the bound state of the kink, which gets unbound at
 $h_1=h_{cr}^{(1)}$. For $h_1>h_{cr}^{(2)}$ such interpretation is impossible
because the energy gap is smaller than the
boundary excitation energy.

Besides the fundamental boundary 1-string, there exists an infinite set of
``long'' boundary strings, consisting of roots $\al_0-2ik\g,
 \al_0-2i(k-1)\g,..., \al_0+2ni\g$ with $n,k\geq 0$. We will call such solution
an (n,k) boundary
 string (thus the fundamental boundary string considered above is the (0,0)
 string). One can use the same arguments as given
in \cite{SS} to show that the (n,k) string is a solution of BE when its
 ``center of mass'' has positive imaginary part
and the lowest root $\al_0-2ik\g$ lies below the real axis.
Thus sufficiently long boundary
string solutions exist even in the region $h^{(1)}_{cr}<h_1<h^{(2)}_{cr}$.
However, a direct calculation shows that their energy vanishes with respect to
the vacuum,
 so they represent charged vacua. \footnote{As an example, consider
the boundary (1,0)-string consisting of the roots $\al_0+2i\g, \al_0$. It
exists
if $-1<H_1<1$, although the (0,0)-string exists only if $-1<H_1<0$.
The (1,0)-string has  charge $S_z=-1$ and vanishing energy with respect to
the vacuum.}
(Analogous phenomenon occurs for the
 ``long''
strings in the bulk \cite{BVV}: if the imaginary part of $\al$
lies outside the strip $-2\g<{\rm Im}\al<2\g$, it gives no contribution
 to the energy.) For $0<h_1<h_{cr}^{(1)}$ and
$h_{cr}^{(2)}<h_1<\infty$ the (n,0) strings also represent charged vacua, while
(n,k) strings with $k\geq 1$ have the same energy
(\ref{ansen}) as the boundary bound state
found above, and
hence represent charged boundary excitations. \footnote{{\it E.g.},
(1,1) string with roots $\al_0+2i\g, \al_0, \al_0-2i\g$ has $S_z=-3/2$
and energy given by (\ref{ansen}) with respect to the physical vacuum.}

Consider now the right boundary, $h_2<0$ ($H_2<-1$). Now the fundamental
boundary 1-string solution $\al_0=-i\g H_2$ exists for any value of $h_2$ in
the
interval $-h_{lim}<h_2<0$
(resp. $\al_0=\pi-i\g H_2$ for $h_2<-h_{lim}$).
Explicit calculation shows that  it has non-vanishing
energy only if $-2<H_2<-1$, which corresponds to $-h_{cr}^{(1)}<h_2<0$ (resp.
$h_2<-h_{cr}^{(2)}$). For such values of $h_2$ the energy  of the 1-string
with respect to the Dirac sea is positive and equal to
$\ve_b(-h_2)$ (see eq. (\ref{ansen})), and its spin is $S_z=-1/2$.
In some sense the pictures are dual for the positive and
negative $h$ cases: there exist two states when $|h|$ is not between
$|h_{cr}^{(1)}|$ and $|h_{cr}^{(2)}|$, one with a boundary 1-string
and one without. One of them is the ground state and another is the excited
state at the boundary, and these states exchange their roles when the sign
of $h$ changes.
 The analysis of long boundary strings is
very similar to that at the left boundary, and therefore  will be
omitted. The net result is again that long boundary strings represent
charged vacua or charged boundary excitations.

In all examples shown above, the charge of boundary excitations turned out to
be half-integer. One can easily check that this is true for all boundary
strings
representing charged excitations.
Since the charge of physical excitations is obviously restricted to be an
 integer (see (\ref{ene})),
we conclude that a boundary excitation can appear only paired with the bulk
excitation of half-integer charge (i.e. containing an odd number of holes),
 or with a boundary excitation at the other end of the chain.
 We give a qualitative interpretation of this fact below.

To compute the vacuum surface energy, eq. (\ref{surfen}), of the
model (\ref{ham}),
 one should use eq. (\ref{ener}) in the limit $L=\infty$ with the root density
determined from eqs. (\ref{inteqII}) or (\ref{inteqI}) and
the boundary terms (\ref{defBI}) or (\ref{defBII}), depending on
the value of $h$.  Define for convenience
\be
g(\Delta)={\Delta\ov 2}+2\sinh\g\sum_{n=1}^{\infty}{e^{-2n\g}-1\ov\cosh
2n\g}.
\label{fung}
\ee
 We  consider separately the following intervals for positive $h_1$
and negative $h_2$:

1) $|h_{1,2}|<h_{cr}^{(1)}$. The ground state contains one boundary 1-string,
corresponding to $h_1$.
 The spin of the ground state can be found to be $S_z=0$.
Using  eqs. (\ref{defBII}),(\ref{pbdry}), and (\ref{inteqI})
and subtracting the bulk contribution (\ref{grener})
we get
\be
E_{surf}={1\ov 2}(h_1+h_2)-g(\Delta)
-\sinh\g\sum_{n=1}^{\infty} (-1)^n{e^{-\g H_1n}-e^{\g H_2n}
\ov \cosh\g n}. \label{eqI}
\ee

2) $|h_{1,2}|>h_{cr}^{(2)}$. The ground state contains one boundary 1-string
and has $S_z=0$.
{}From  eqs. (\ref{defBI}) and (\ref{inteqI}) it follows:
\be
E_{surf}={1\ov 2}(h_1+h_2)-g(\Delta)
-\sinh\g\sum_{n=1}^{\infty} {e^{-\g H_1n}-e^{\g H_2n}
\ov \cosh\g n}. \label{eqII}
\ee

3)  $h_{cr}^{(1)}<|h_{1,2}|<h_{lim}$. The ground state has no boundary strings
and its spin is zero.
{}From (\ref{inteqII}) and (\ref{defBII}) one obtains
the same expression as in case 1).

4) $h_{lim}<|h_{1,2}|<h_{cr}^{(2)}$. From (\ref{inteqII}) and (\ref{defBI}) one
 obtains the same expression as in case 2). The ground state has the same
 structure as in case 3).

A qualitative plot of the surface energy as a function of $h$  ($h=h_1=-h_2$)
is given in Fig.1.  The apparent difference between (\ref{eqI}) and
(\ref{eqII})
 is an artefact of our parametrization of $h$ in terms of
$H$. In fact, $E_{surf}$ is an analytic function of $h$ in the domain
 $h\in(0,\infty)$, which can be seen after substituting $H$ as a function of
$h$  according to (\ref{maphI})-(\ref{maphII}).
In this sense the fields $h_{cr}^{(1,2)}$ are not actually ``critical.''
We find for $h_1=h_2=0$ the value

\be
E_{surf}=-{\Delta \ov 2}+4\sinh\g\left( {1 \ov 4}+\sum_{n=1}^{\infty}{\frac
{e^{2n\g}-1}{1+e^{4n\g}}}+
\sum_{n=1}^{\infty}{\frac{(-1)^n}{1+e^{2n\g}}}\right).
\label{Esurf}
\ee

In the extreme anisotropic limit $\Delta\to\infty$, $h\sim\Delta$ of the $XXZ$
chain (\ref{ham}) one gets the one-dimensional Ising model:
\be
{\cal H}= {1\over 2 }
\left\{\sum_{i=1}^{L-1}
\Delta\sigma^z_i
\sigma^z_{i+1} +h_1\sigma_1^z+h_2\sigma_L^z \right\}, \label{hamising}
\ee
In this limit from (\ref{maphI})-(\ref{maphII}) one has
\be
h\approx \Delta \pm e^{-\g H}, \label{limith}
\ee
and the gap between $h_{cr}^{(1)}$ and  $h_{cr}^{(2)}$ dissapears, so
for any $h$ there
exists a boundary bound state. The energy of the ``bulk'' hole (\ref{hole})
 becomes
independent of $\theta$ and equal to $\Delta$,
since only $n=0$ term contributes to the sum. The energy of the boundary bound
state (\ref{ansen})
becomes $\ve_b=\Delta \pm e^{-\g H}=
h$. This suggests the following interpretation
in terms of the Ising chain: in the Ising ground state the $i$-th spin
has the value $(-1)^i$. Local bulk excitation of the smallest energy $2\Delta$
can be obtained by flipping one spin
(the first spin excepted). The arising two surfaces (domain walls) separating
 the flipped spin  from its right and left neighbours
are called kinks and carry the energy $\Delta$ each. The kink corresponds
to a hole in the Dirac sea in the Bethe ansatz picture, and these holes can
 exist only in pairs,
which is obviously the case for the kinks in the Ising chain.  The charge of
the
one-spin-flipped state is equal to one, in agreement with the charge of
two holes in BA. In addition to charge 1 excitation, one has charge 0
excitation of the same energy obtained by flipping any even number of spins
in a row. In the BA this corresponds to the ``2 holes and 2-string'' state.
The boundary bound state is obtained by flipping the first (or the last)
spin. Such a
state has the energy $h_1+\Delta$ above the vacuum energy in the
Ising limit,
where $h_1$ is the contribution of the boundary term in eq. (\ref{hamising})
and $\Delta$ is the energy of the kink created because of the
interaction of the boundary spin with the bulk ones. Thus flipping
the boundary spin actually gives
a combination of the boundary excitation and the bulk kink.
Still another possibility is to flip all spins, creating two boundary
bound states, one at each boundary.
 This explains why, in the BA picture, a boundary excitation
can exist only if paired with a hole in the Dirac sea or with another boundary
excitation.

 The ground state surface energy (\ref{surfen})
of the Ising chain in the thermodynamic limit is $(\Delta-h_1+h_2)/2$.
The $\Delta/2$ contribution here is the bulk interaction energy that we
lost when we
disconnected the periodic chain, and $\pm h_{1,2}/2$ is the contribution of
each
of the boundary terms. Taking the limit $\g\to\infty$ in eqs. (\ref{eqI})-
(\ref{eqII}), we obtain the expected result
 $E_{surf}\to(\Delta-h_1+h_2)/2$.

In the isotropic (rational)
limit $\Delta\to 1$ ({\it i.e.}, $\g\to 0$) one gets
the XXX chain in a boundary magnetic field,
which was discussed in the BA framework in \cite{GMN} for $0<h_1,h_2<2$. From
(\ref{maphI})-(\ref{maphII}) one has in this limit
\be
h={2\ov 1+H}. \label{Hnew}
\ee
There is only one critical field $h_{cr}=2$, which is the limit
of $h_{cr}^{(2)}$. Passing from summation to integration
in eq. (\ref{eqII}), we obtain for $0<h_1,-h_2<\infty$:
\begin{eqnarray}
E_{surf}&=&{1\ov 2}(h_1+h_2)-{1\ov 2} +{\pi\ov 2} -\int_{0}^{\infty}dx
{e^{-({2\ov h_1}-1)x} - e^{({2\ov h_2}-1)x}+ e^{-x}
\ov\cosh x} \nonumber \\
&=&{1\ov 2}(h_1-h_2)-{1\ov 2} +{\pi\ov 2} -\int_{0}^{\infty}dx
{e^{-({2\ov h_1}-1)x} + e^{({2\ov h_2}+1)x}+ e^{-x}
\ov\cosh x},
\label{xxxlimI}
\end{eqnarray}
where the second line was obtained from the first one after a simple
manipulation. We can compare this result with that of
 \cite{GMN}
and find the agreement. For $h_1=h_2=0$
one has from (\ref{xxxlimI}):
$E_{surf}=(\pi-1)/2 - \ln 2$.

Another aspect is the structure of the ground state
in the regime $h_{1,2}>0$. Assuming that, for example, for
$h_{1,2}>h_{cr}^{(2)}
$ the ground state contains both left and right boundary 1-strings
to minimize the energy, we
end up after a short calculation with a half-integer spin of the vacuum,
which signals that such a state cannot,
in fact, be the vacuum. Hence, the ground state must
have a more intricate structure.
Appealing to the Ising limit $\g\to\infty$, one
sees that for $h_{1,2}>\Delta$ the ground state
 must have both boundary spins  directed opposite to the magnetic field,
 and therefore contain a kink in the bulk (recall that L is even).
On the other hand, for $h_{1,2}<\Delta$ the lowest energy configuration
is such that the boundary spins are antiparallel, which means that the
 physical vacuum contains what was
called a boundary excitation at one of the ends. This suggests that for finite
$\Delta$ the correct ground state
 wave-function of the Hamiltonian (\ref{ham}) should contain a bulk hole
with the minimal possible energy (i.e. the kink with zero rapidity
$\theta=0$) and both boundary 1-strings
 when $h_{1_2}>h_{cr}^{(2)}$.  Such a state has spin zero.
Since the energy difference between this ground state and a state with a moving
(instead of statinary)
kink vanishes when
the rapidity of the kink goes to zero, the spectrum acquires
a gapless branch.
\footnote{In the Ising limit $\g\to\infty$ the energy of the kink is
independent
of rapidity, and therefore this branch degenerates to the vacuum.}
 Similarly, when $h_{cr}^{(1)}<h_{1,2}<h_{cr}^{(2)}$, for the
ground state
to have the integer charge, it should also contain a kink in the bulk. When
$h_{1,2}<h_{cr}^{(1)}$ the physical vacuum contains only
one of the two boundary 1-strings and no stationary kink in the bulk
(when $h_1=h_2$ there are two possibilities to have either left or right
boundary 1-string in the vacuum, corresponding to the obvious two-fold
degeneracy of the Ising ground state in this case).
  Such a state has a smaller energy for $h_{1,2}<h_{cr}^{(1)}$ than the one
with
 a hole in the bulk and two boundary strings,
while for $h_{1,2}>h_{cr}^{(2)}$ the state with the bulk hole is energetically
 preferable, since in this case $\ve_b>\ve_h$ (see Fig.2 and \cite{MJ}).
This situation is in
some sense analogous to the case of the periodic $XXZ$ chain with odd L,
where the ground state contains a kink.
Let us mention that
according to the above discussion
 the surface energy in the case $h_{1,2}>h_{lim}$
is:
\be
E_{surf}={1\ov 2}(h_1+h_2)-g(\Delta)+\varepsilon_h(0)
-\sinh\g\left(1+\sum_{n=1}^{\infty} {e^{-\g H_1n}+e^{-\g H_2n}
\ov \cosh\g n}\right). \label{eqM}
\ee
In the rational ($\g\to 0$) limit $\varepsilon_h(0)$ vanishes
and eq. (\ref{eqM}) becomes
\be
E_{surf}={1\ov 2}(h_1+h_2)-{1\ov 2} +{\pi\ov 2} -\int_{0}^{\infty}dx
{e^{-({2\ov h_1}-1)x} + e^{-({2\ov h_2}-1)x}+ e^{-x}
\ov\cosh x}. \label{xxxlimIP}
\ee
This expression agrees with the one obtained in \cite{GMN}. Note that
the authors of
\cite{GMN} obtained eq. (\ref{xxxlimIP}) under assumption that
$0<h_{1,2}<h_{cr}$, while our derivation shows that this result is valid for
$0<h_{1,2}<\infty$. In the Ising limit eq. (\ref{eqM}) gives the correct
result $E_{surf}=(3\Delta-h_1-h_2)/2$.

 In conclusion we would like to mention that within the BA technique
it is possible to calculate also the boundary S-matrix for the
scattering
of kinks (represented by holes in the Dirac sea) in the ground state of the
Hamiltonian (\ref{ham}) or in
the excited boundary state.
Such a calculation has been performed
 in \cite{GMN} for the boundary XXX chain
and in \cite{SS,FS} for the boundary sine-Gordon model.
For the boundary XXZ chain these
S-matrices have been obtained by Jimbo et al.,\cite{MJ} by the algebraic
approach.

\acknowledgements
One of the authors (S.S.) would like to thank H.Saleur
for helpful discussions.

\begin{figure}
\epsfxsize=18truecm
\centerline{\epsfbox{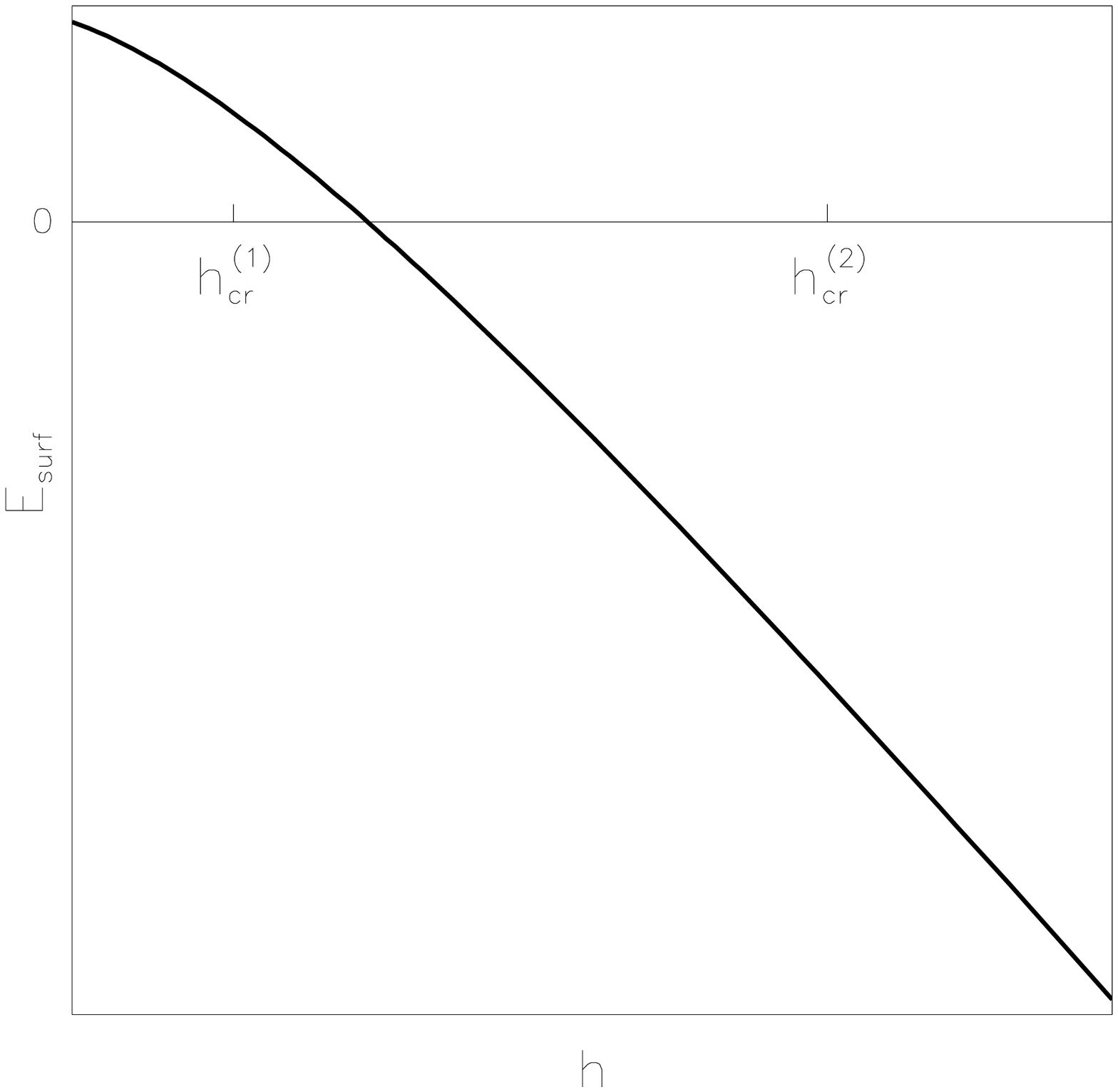}}
\caption{A schematic plot of the ground state surface energy as a function of
 the boundary magnetic field $h=h_1=-h_2$. The critical fields are also shown.}
\end{figure}

\begin{figure}
\epsfxsize=18truecm
\centerline{\epsfbox{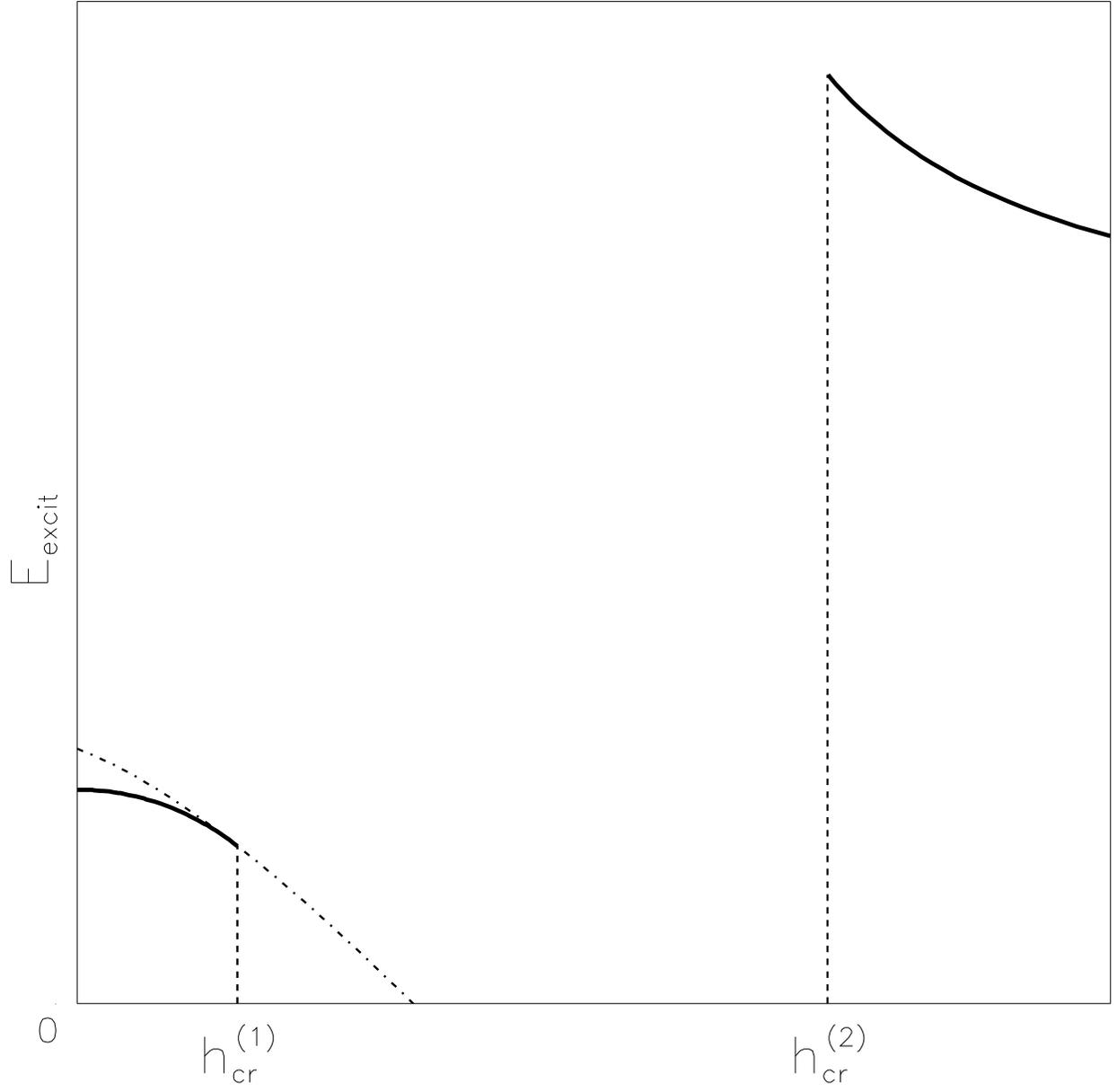}}
\caption{ Solid line: a schematic plot of the surface energy of the state with
a
 boundary excitation at one of the ends, $E_{surf}(h)+\ve_b(h)$, as a function
 of
 the boundary magnetic field $h=h_1=-h_2$. Dash-dotted line: the energy of
the state containing a zero rapidity hole in the bulk $E_{surf}(h)+\ve_h(0)$.}
\end{figure}

 \end{document}